\documentclass{emulateapj}

\newcommand{\kms}{\mbox{km s$^{-1}~$}} 
\newcommand{\kmse}{\mbox{km s$^{-1}$}}

\newcommand{\vgsr}{$V_{\rm GSR}~$}
\newcommand{\vgsre}{$V_{\rm GSR}$}
\newcommand{\dgr}{$^{\circ}~$}

\newcommand{\hi}{\ion{H}{1} }

\begin{document}

\title{The Apache Point Observatory Galactic Evolution Experiment: \\
First Detection of High Velocity Milky Way Bar Stars}

\shorttitle{APOGEE Bar Kinematics}
\shortauthors{NIDEVER ET AL.}

\author{David L. Nidever\altaffilmark{1},
Gail Zasowski\altaffilmark{1},
Steven R. Majewski\altaffilmark{1},
Jonathan Bird\altaffilmark{2},
Annie C. Robin\altaffilmark{3},
Inma Martinez-Valpuesta\altaffilmark{4},
Rachael L. Beaton\altaffilmark{1},
Ralph Sch\"onrich\altaffilmark{2},
Mathias Schultheis\altaffilmark{3},
John C. Wilson\altaffilmark{1},
Michael F. Skrutskie\altaffilmark{1},
Robert W. O'Connell\altaffilmark{1},
Matthew Shetrone\altaffilmark{5},
Ricardo P. Schiavon\altaffilmark{8},
Jennifer A. Johnson\altaffilmark{2},
Benjamin Weiner\altaffilmark{9},
Ortwin Gerhard\altaffilmark{4},
Donald P. Schneider\altaffilmark{10},
Carlos Allende Prieto\altaffilmark{11},
Kris Sellgren\altaffilmark{2},
Dmitry Bizyaev\altaffilmark{7},
Howard Brewington\altaffilmark{7},
Jon Brinkmann\altaffilmark{7},
Daniel J. Eisenstein\altaffilmark{12},
Peter M. Frinchaboy\altaffilmark{13},
Ana Elia Garc{\'{\i}}a P{\'e}rez\altaffilmark{1},
Jon Holtzman\altaffilmark{6},
Fred R. Hearty\altaffilmark{1},
Elena Malanushenko\altaffilmark{7},
Viktor Malanushenko\altaffilmark{7},
Demitri Muna\altaffilmark{14},
Daniel Oravetz\altaffilmark{7},
Kaike Pan\altaffilmark{7},
Audrey Simmons\altaffilmark{7},
Stephanie Snedden\altaffilmark{7},
Benjamin A. Weaver\altaffilmark{14}
}

\altaffiltext{1}{Department of Astronomy, University of Virginia,
Charlottesville, VA, 22904-4325, USA (dnidever@virginia.edu)}

\altaffiltext{2}{Department of Astronomy and the Center for Cosmology and
Astro-Particle Physics, The Ohio State University, Columbus, OH 43210, USA}

\altaffiltext{3}{Institut Utinam, CNRS UMR 6213, OSU THETA, Universit{\'e} de Franche-Comt{\'e},
41bis avenue de l'Observatoire, 25000 Besan\c{c}on, France}

\altaffiltext{4}{Max-Planck-Institut f\"ur Extraterrestrische Physik, Giessenbachstrasse,
85748 Garching, Germany}

\altaffiltext{5}{University of Texas at Austin, McDonald Observatory, Fort Davis, TX 79734, USA}

\altaffiltext{6}{New Mexico State University, Las Cruces, NM 88003, USA}

\altaffiltext{7}{Apache Point Observatory, Sunspot, NM 88349, USA}

\altaffiltext{8}{Gemini Observatory, 670 N. A'Ohoku Place, Hilo, HI 96720, USA}

\altaffiltext{9}{Steward Observatory, 933 N. Cherry St., University of Arizona, Tucson, AZ 85721, USA}

\altaffiltext{10}{Department of Astronomy and Astrophysics, The Pennsylvania State University, University Park, PA 16802, USA}

\altaffiltext{11}{Instituto de Astrof\'{i}sica de Canarias, E38205 La Laguna, Tenerife, Spain}

\altaffiltext{12}{Harvard-Smithsonian Center for Astrophysics, 60 Garden St., MS \#20, Cambridge, MA 02138}

\altaffiltext{13}{Texas Christian University, Fort Worth, TX 76129, USA}

\altaffiltext{14}{Center for Cosmology and Particle Physics, New York University, New York, NY 10003 USA}

\begin{abstract}
Commissioning observations with the Apache Point Observatory Galactic Evolution Experiment (APOGEE),
part of the Sloan Digital Sky Survey III, have produced radial velocities (RVs) for $\sim$4700 K/M-giant
stars in the Milky Way bulge.
These high-resolution ($R\sim22,500$), high-$S/N$ ($>$100 per resolution element),
near-infrared (1.51-1.70 $\mu$m; NIR) spectra provide accurate RVs ($\epsilon_{\rm V}$$\sim$0.2 \kmse)
for the sample of stars in 18 Galactic bulge fields spanning
$-1$\degr$<l<20$\degr, $|b|<20$\degr, and $\delta>-32$\degr.
This represents the largest NIR high-resolution spectroscopic sample of giant stars ever assembled
in this region of the Galaxy.
A cold ($\sigma_{\rm V}$$\sim$30 \kmse), high-velocity peak (\vgsre$\approx$+200 \kmse) is found to comprise
a significant fraction ($\sim$10\%) of stars in many of these fields.
These high RVs have not been detected in previous MW surveys and are not
expected for a simple, circularly rotating disk.
Preliminary distance estimates rule out an origin from the background Sagittarius tidal stream
or a new stream in the MW disk. 
Comparison to various Galactic models suggests that
these high RVs are best explained by stars in orbits of the Galactic bar potential, although
some observational features remain unexplained.
\end{abstract}


\section{Introduction}
\label{sec:intro}

Multiple $N$-body models have demonstrated that rotating stellar
disks form bars relatively quickly under a wide variety of conditions
\citep[e.g.,][]{Bureau05,Debattista06,Martinez-Valpuesta06},
and observations have identified a high incidence of 
large central bars in extragalactic systems \citep[e.g.,][]{Lutticke00,Chung04}.
Even in edge-on spiral galaxies, where the stellar bar can be difficult to isolate 
photometrically from the surrounding bulge and/or disk,
the bar can be identified
via kinematical signatures: double-peaked rotation curves, radial velocity (RV) dispersion profiles
with particular combinations of plateaus and peaks/minima, and the relation between
the mean and higher moments of the RV distribution \citep[e.g.,][]{Bureau05}.

In recent decades, good evidence for a Milky Way (MW) bar has arisen from infrared (IR)
surface brightness maps and star counts \citep[e.g.,][]{Blitz91,Weinberg92,Dwek95,Hammersley00,Cole02,Benjamin05,Lopez-Corredoira07,Robin12},
carbon stars \citep[][]{Cole02}, the solar neighborhood velocity field \citep[e.g.,][]{Dehnen00,Minchev07},
microlensing \citep[e.g.,][]{Gyuk99,Rattenbury07},
and inner Galaxy {\it gas} kinematics \citep[e.g.,][]{Liszt80,Binney91,Fux99,Weiner99}.
However, the high extinction towards the inner Galaxy midplane ($A_V\approx$5--10) 
has long precluded an extensive spectroscopic survey of the type required to measure the bulge/bar 
{\it stellar} kinematics on par with studies of extragalatic bars in galaxies viewed edge-on.

\begin{figure*}[ht!]
\begin{center}
$\begin{array}{cc}
\includegraphics[angle=0,scale=0.35]{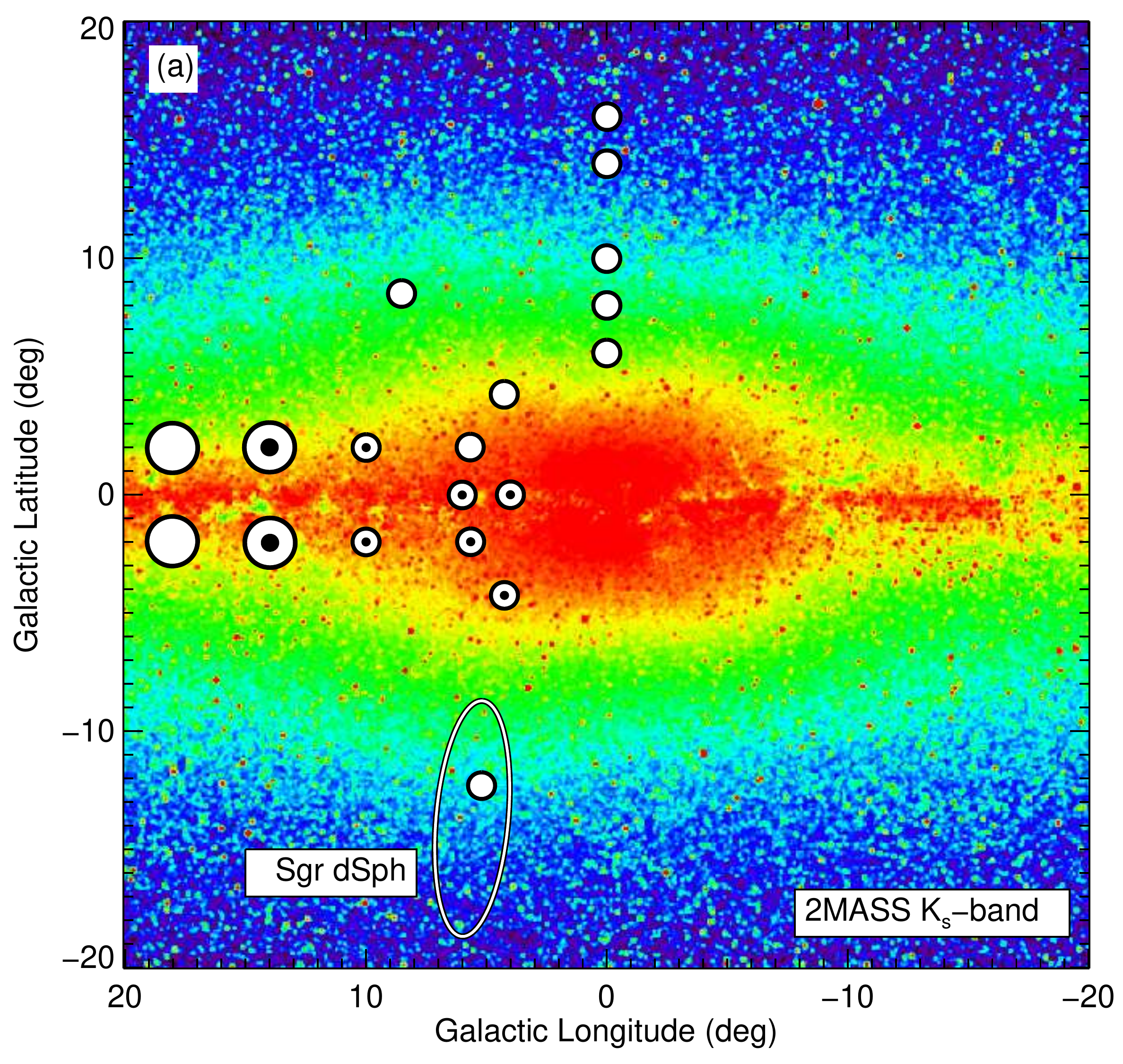}
\includegraphics[angle=0,scale=0.45]{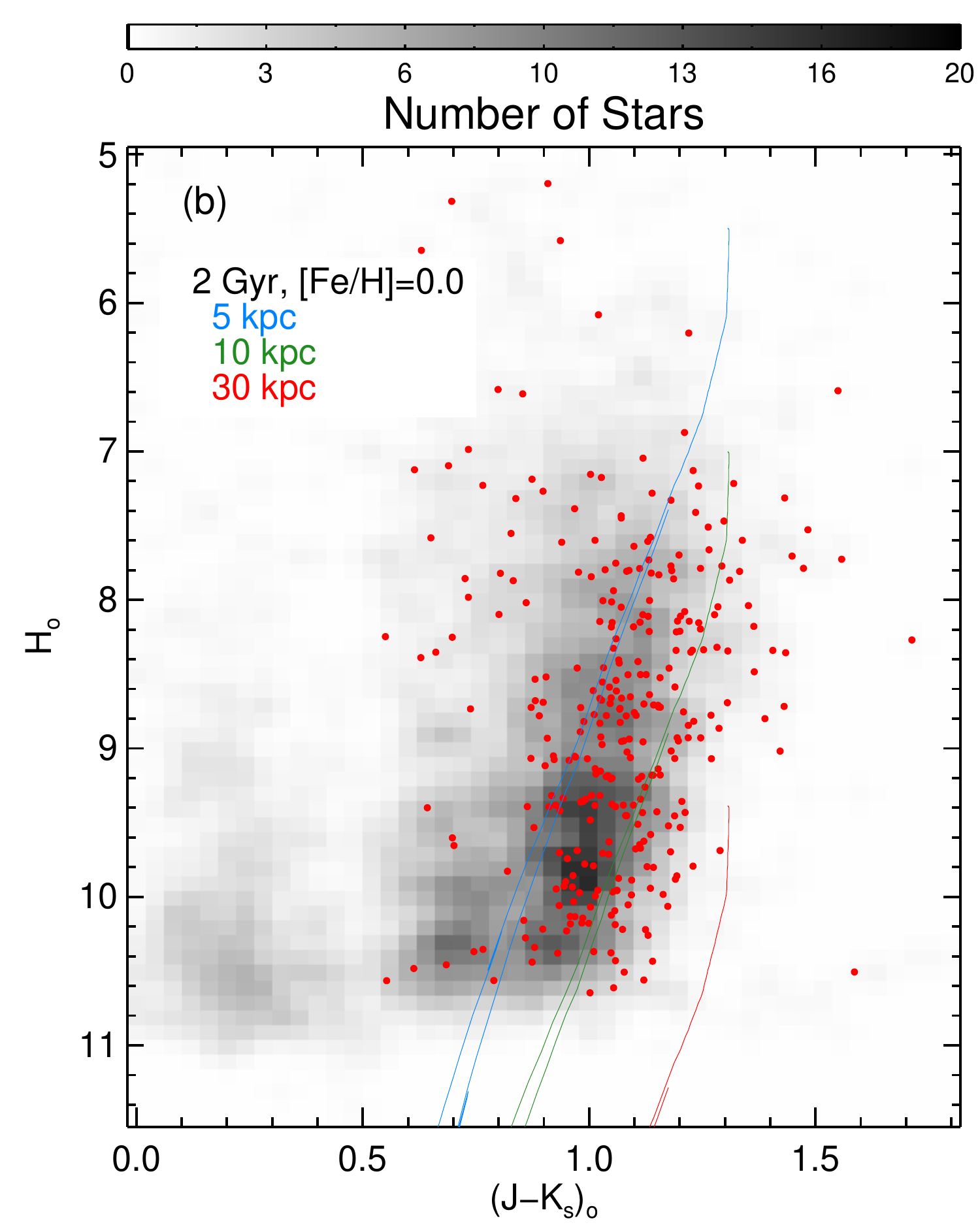}
\end{array}$
\end{center}
\caption{(a) Coverage of APOGEE commissioning fields (white circles) in the Galactic bulge and inner disk regions.
The eight fields with distinct high-velocity peaks are marked with black dots.
The background image is the 2MASS $K_{\rm s}$ band \citep{Skrutskie06} showing the boxy bulge. The
Sagittarius dwarf spheroidal is indicated by an ellipse.
(b) Reddening-corrected Hess diagram of all observed stars (grayscale) with
the high-velocity stars with $[J-K_{\rm s}]_0\ge0.5$
shown as red dots.  The high-velocity thresholds
in Table \ref{table_rv}, which represent the RV of the ``trough'' between the main peak and the high-velocity
peak, were used to select the high-velocity stars from each field.
Fiducial \citet{Girardi02} isochrones (2 Gyr, [Fe/H]=0.0) are shown at distances of 5/10/30 kpc
(blue/green/red).}
\label{fig_mapcmd}
\end{figure*}

More recently, the BRAVA \citep{Rich07b} and ARGOS \citep{Ness12} surveys
have begun performing kinematical measurements in the bulge periphery, mapping the bulk
cylindrical (i.e., bar-like) rotation of the bulge.  
In this Letter, we report the discovery, within data from the new APOGEE
survey, of a ``cold'' kinematical feature in the inner disk, which has
not been seen in previous surveys and is, most likely, a signature of
particular stellar orbits in the MW bar.
In \S\ref{sec:red}, we briefly describe the observations
and data reduction, and we present in \S\ref{sec:rvhist} the RV distributions of the 
APOGEE fields, along with the RV predictions of multiple Galactic kinematical models.
In \S\ref{sec:discussion}, we use these comparisons to interpret the nature of 
the APOGEE observations, and we evaluate alternative explanations for the data and model discrepancies.

\section{Observations And Data Reduction}
\label{sec:red}

The APOGEE project \citep[][Majewski et al.\ in preparation]{Eisenstein11}
uses a custom-built, cryogenic spectrograph \citep{Wilson10} recording 300 simultaneous
near-IR spectra (1.51--1.70$\mu$m) fed
from the Sloan 2.5-m telescope \citep{Gunn06}.
The final configuration for the spectrograph is described in Wilson et al.\ (in preparation);
however, the data described here were taken before the instrument was in 
its fully commissioned state and in optimal focus (achieved September 2011).
The net effect was a somewhat blurred line-spread-function, yielding a
degradation of the resolution primarily in the red detector to $R\sim$16,000 (1.65--1.70$\mu$m)
compared to $R\sim23,000$ in the other two detectors (1.51--1.58$\mu$m and 1.59--1.64$\mu$m). 

The APOGEE fibers are plugged into standard Sloan 2.5-m plugplates and
observed similarly to those of
the optical Sloan spectrographs, with the following 
variations: (1) 35 fibers in each plugplate configuration collect sky spectra, 
(2) another 35 fibers are placed on bright, hot 
stars to gauge telluric (CO$_2$, H$_2$O, and CH$_4$) absorption, 
and (3) because the large zenith distances required to observe the bulge from APO 
cause strong differential refraction, 
the field diameter was restricted to 1\dgr for $l$$\lesssim$12\dgr fields (2\dgr for the rest).

We observed 18 bulge fields in June--July 2011 with $-1$\degr$<l<20$\degr, $|b|<20$\dgr and $\delta>-32$\dgr
(see Fig.\ \ref{fig_mapcmd}a).
Targets were selected from the 2MASS Point Source Catalog \citep{Skrutskie06} requiring $(J-K_{\rm s})_0\ge0.5$ 
(although some plugplates had no color selection for testing purposes)
and $H$$\leq$11.0 (see Zasowski et al., in preparation, for details on APOGEE
targeting strategy).\footnote{The 2MASS photometry was dereddened using the RJCE method 
\citep{Majewski11}, with WISE \citep{Wright10} and GLIMPSE \citep{Churchwell09}
providing the required mid-IR photometry.}
The dereddened Hess diagram for all observed stars is shown in Figure \ref{fig_mapcmd}b.
Each field was observed for a total integration of roughly one hour,
for a $S/N$$>$100 per APOGEE pixel for each star.

\begin{figure*}[ht!]
\begin{center}
\includegraphics[angle=0.0,scale=0.7]{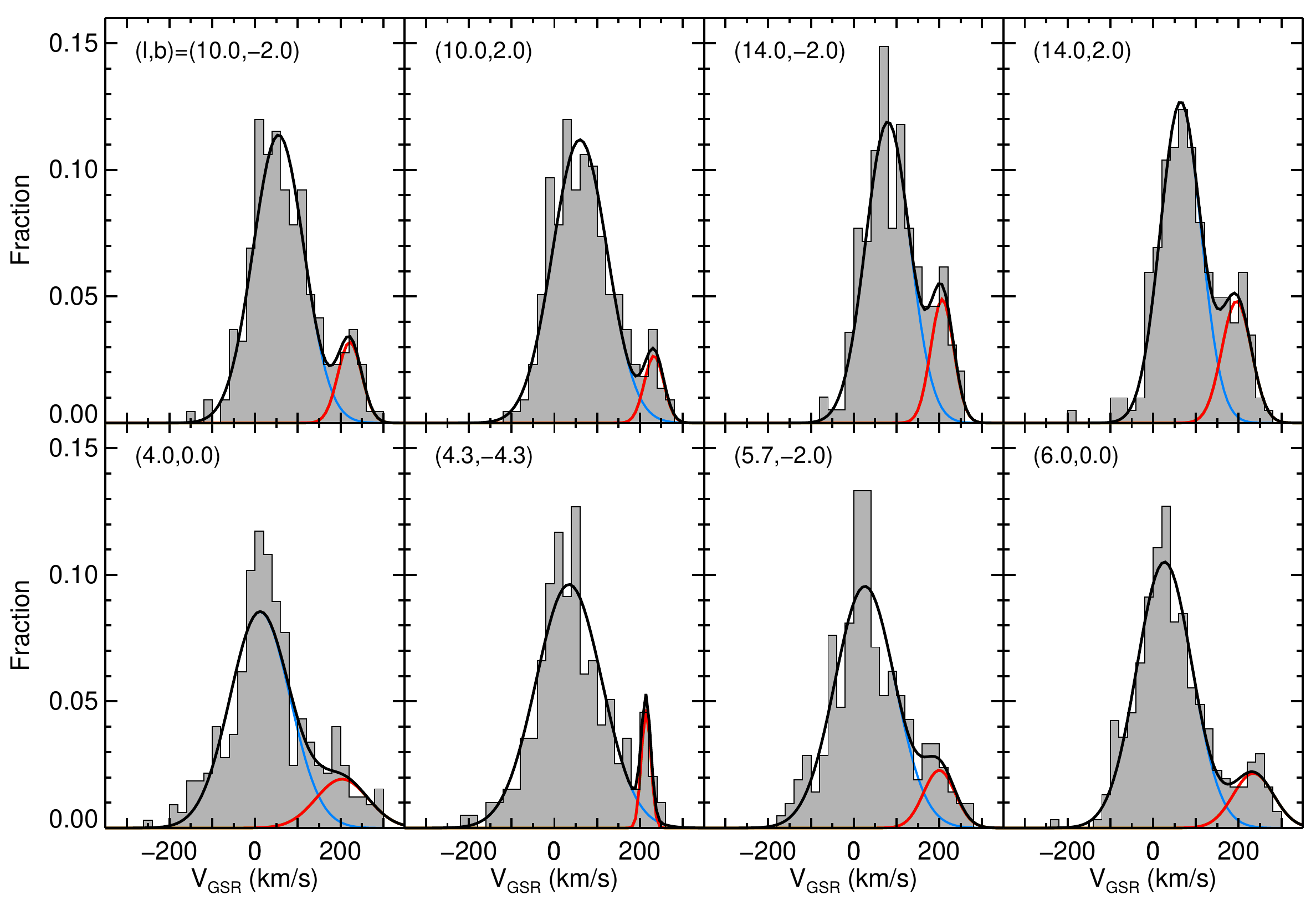}
\end{center}
\caption{Velocity ($V_{\rm GSR}$) histograms of eight APOGEE bulge fields
showing a dual-peak structure, using 20 \kms binning.
A two-Gaussian model (see text) is fitted to the data to determine the central velocity of both peaks
(blue/red lines).
The solid black line is the sum of the two components.
}
\label{fig_rvhist2}
\end{figure*}

An automated data reduction pipeline written specifically for APOGEE data (Nidever et al., in 
preparation) was used for (1) collapsing the sample-up-the-ramp datacubes to 2D images with rejection of cosmic rays,
(2) extraction of 1D spectra from the 2D images, (3) wavelength calibration using
ThArNe/UrNe arc lamp exposures and airglow lines, 
(4) correction of telluric absorption as monitored by the 35 hot star standards, 
(5) subtraction of airglow lines and sky continuum, and (6) derivation of RVs
by cross-correlation against a grid of template spectra spanning the expected range of temperatures,
gravities, and metallicities of target stars.
The median RV scatter for stars observed multiple times is 0.22 \kms (with a mode of $\sim$0.09 \kmse),
while tests of derived RVs
for observations of 53 stars in three
globular clusters (M3, M13, M15)
with accurate literature RVs suggest that the zeropoint of the APOGEE pipeline RVs is accurate
to $\sim$0.26$\pm$0.22 km s$^{-1}$.

\section{Radial Velocity Distributions: Data and Models}
\label{sec:rvhist}

The bulge RV distributions exhibit a significant (and unexpected) number of high-velocity stars, which
appear as prominent high-velocity peaks in eight fields and as high-velocity ``shoulders''
to the main peak in many of the other fields.
We focus on the RV distributions of the eight APOGEE bulge fields with distinct high-velocity peaks,
shown in Figure \ref{fig_rvhist2} with a $(J-K_{\rm s})_0\ge0.5$ selection.
The stars in the high-RV peak account for $\sim$10\% of the stars in a field.
To derive the mean velocity of the high-velocity component, the histograms are fitted with a
two-Gaussian model -- a hot component for the bulge and a colder component for the high-velocity
stars.\footnote{The models are not intended to be
perfect ``physical'' representations and are poor fits for several fields (e.g., [4.0,0.0] and [5.7,-2.0]),
but they serve well to estimate the peak center.}

\begin{figure*}[ht!]
\begin{center}
\includegraphics[angle=0,scale=0.39]{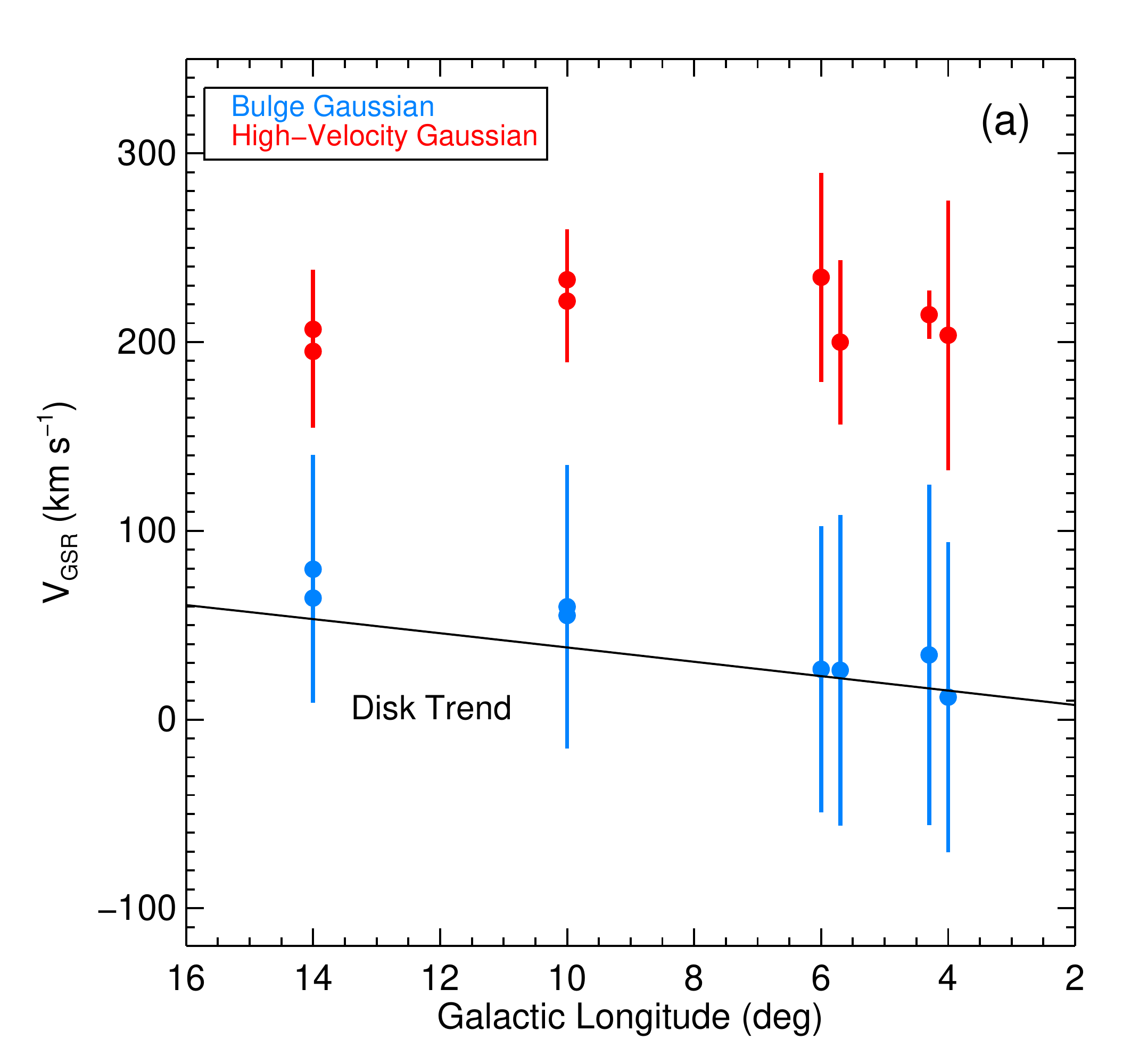}
\includegraphics[angle=0,scale=0.37]{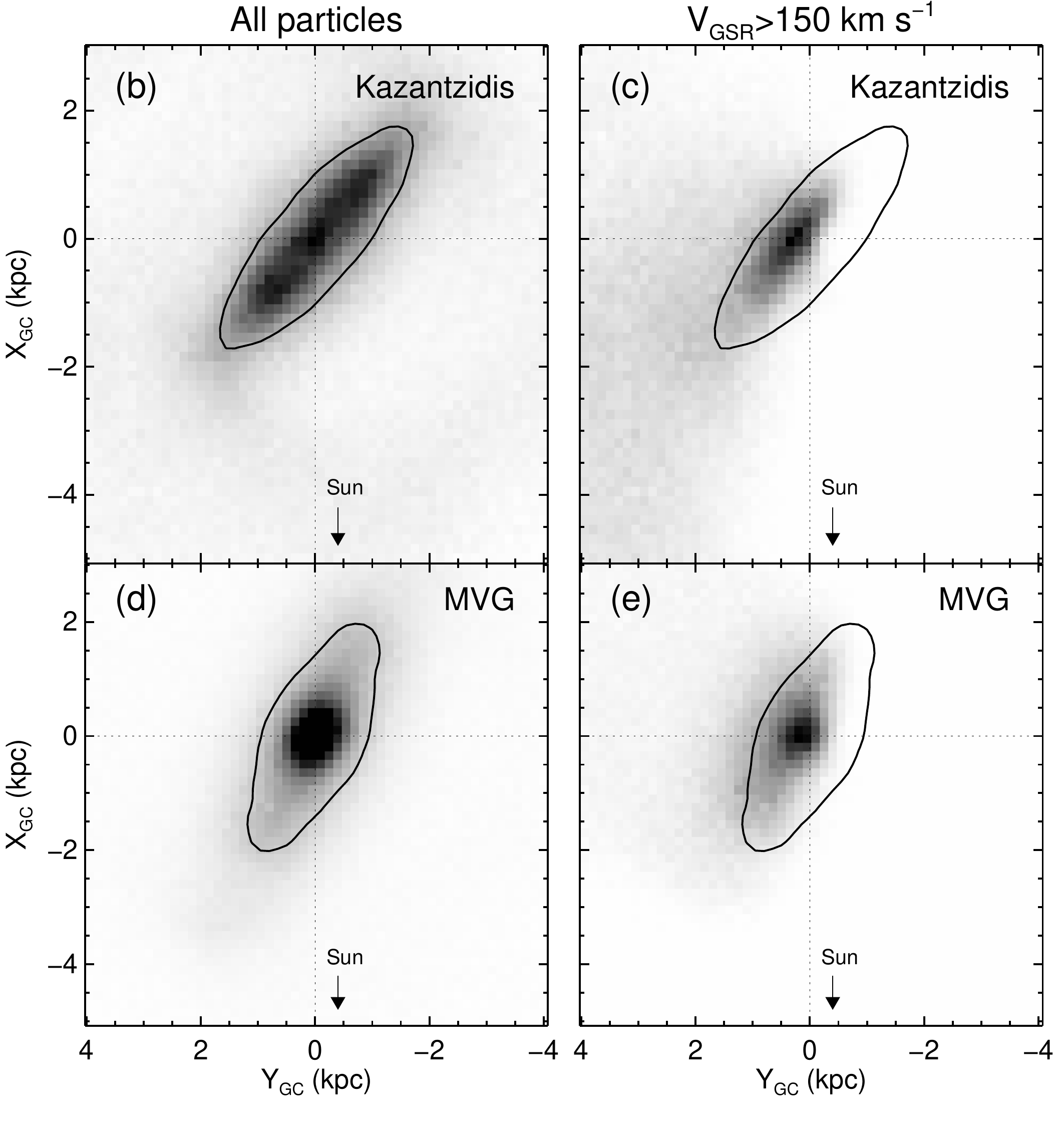}
\end{center}
\caption{(a) Position-velocity diagram of the two-component Gaussian fits to the
APOGEE bulge RV histograms, with main peaks in blue and high-velocity peaks in red.
Gaussian FWHM values are indicated by the vertical lines.
A general trend is seen in the high-velocity component, with mean velocity first rising
with longitude and then slowly falling at longitudes above $l$$\approx$6--7\degr.
The mean velocities of the main components fall close to the 
theoretical disk trend calculated from a simple rotation curve (\vgsre=220 $\sin{l}$; black).
(b--e) Spatial distribution of particles in the Kazantzidis 200 pc scale-height model (upper panels, b+c)
and the MVG model (lower panels, d+e).
All particles are shown in the left panels (b+d) while the right panels (c+e) show the distribution of
particles with \vgsr$>$150 \kms which live on the leading edge of the bar.
The sun's position is at (0,-8.5) which is off the lower edge of the panels.
The black contours outline the general shape of the model MW bar.
}
\label{fig_simrvtrend}
\end{figure*}

These high-velocity peaks have not been previously detected in the MW bulge, including in the
extensive BRAVA data \citep[$-10$\degr$<l<+10$\degr, $b=-4$\degr,$-6$\degr,$-8$\degr;][]{Kunder12}.
Several narrow peaks in the RV histogram were detected in the BRAVA data \citep{Howard08} but were
eventually determined to be due to stochastic effects.  Our high-velocity peaks are unlikely to be due to
stochastic effects because: (1) our sample per field is large (twice that of BRAVA),
(2) the high-velocity peaks are seen in at least eight APOGEE fields,
(3) the peaks appear at similar RVs following a general trend with longitude (Fig.\ \ref{fig_simrvtrend}a),
and (4) the pattern is ``mirrored'' across the midplane 
(e.g., strong similarities between $l$,$b$=10\degr,$\pm$2\degr).
We conclude that the high-velocity peaks are due to real motions of stars in the Galaxy.

Table \ref{table_rv} gives the field name, longitude, latitude, diameter, mean \vgsr and velocity dispersion of
all stars, number of stars, mean \vgsr of the high-velocity stars, the high-velocity threshold (essentially the
velocity of the ``trough'' between the main and high-velocity peaks), and the number of high-velocity stars
in the eight APOGEE bulge fields that exhibit the most distinct high-velocity peaks.
To ascertain the nature of the high-velocity stars, we compare the APOGEE RV histograms to several models.

\subsection{Besan\c{c}on Galaxy Model}
\label{subsec:Besancon}

The Besan\c{c}on Galaxy Model \citep[][hereafter BGM]{Robin03, Robin12}
simulates the stellar content in any given line of sight and computes the
photometry, kinematics, and metallicity of each simulated star.  For each
population (thin disk, thick disk, halo, bar, and bulge), a star formation rate history, age, and initial mass function are
assumed, allowing the model to generate the distribution function of absolute magnitude,
effective temperature, and age of the stars. Density functions are assumed for each
population and tested against observations using photometric star counts.
The model also includes a 3D extinction map \citep{Marshall06} and simulated uncertainties
on the observational parameters of each star. 

The BGM bar and bulge populations are described in \citet{Robin12}.
The bar is the more massive component and dominates
the stellar content at low latitudes, while the bulge is longer and thicker and gives a
contribution at intermediate latitudes where the bar becomes less prominent.
The bar kinematics are taken from the \cite{Fux99} dynamical model, and the
bulge kinematics are established to reproduce the 
BRAVA survey data  (Robin et al., in preparation).
This multi-population model also explains well the presence of double
red clumps at medium latitudes \citep{Nataf10,McWilliam10,Saito11}
that are created by stars trapped in vertical resonances associated with the
bar \citep[e.g.,][]{Combes90}, lifting them to high altitudes.

\subsection{Kazantzidis N-body Model}
\label{subsec:Kazantzidis}

We also compare to the $N$-body model of
\citet{Kazantzidis08}, which represents a MW--sized galaxy with
initial disk scale-height of 200 pc.
The model was subjected to a
cosmologically-motivated satellite accretion history and formed a bar
as a consequence of the accretion events. The properties of the bar
are similar to those of the MW bar (in overall orientation and size).

\subsection{Martinez-Valpuesta \& Gerhard Bar Model}
\label{subsec:MVG}

The model developed in \citet[][hereafter MVG]{Martinez-Valpuesta11} is an $N$-body galaxy
based on secular evolution of disk galaxies through angular momentum
transfer between disk and halo. Once the bar appears, it grows stronger
and buckles, forming a boxy bulge, such as the one seen in the
MW.  After this buckling the bar weakens and resumes growth
by giving angular momentum to the live dark matter halo. At this point
the bar has a short, thick component and longer, thinner component.
This model has already been quantitatively compared with
the innermost region of the MW
\citep[$-8$\degr$<l<8$\degr;][]{Gerhard12} and also in the outer bar
\citep{Martinez-Valpuesta11} with good agreement.

\begin{figure*}[ht!]
\begin{center}
\includegraphics[angle=0,scale=0.80]{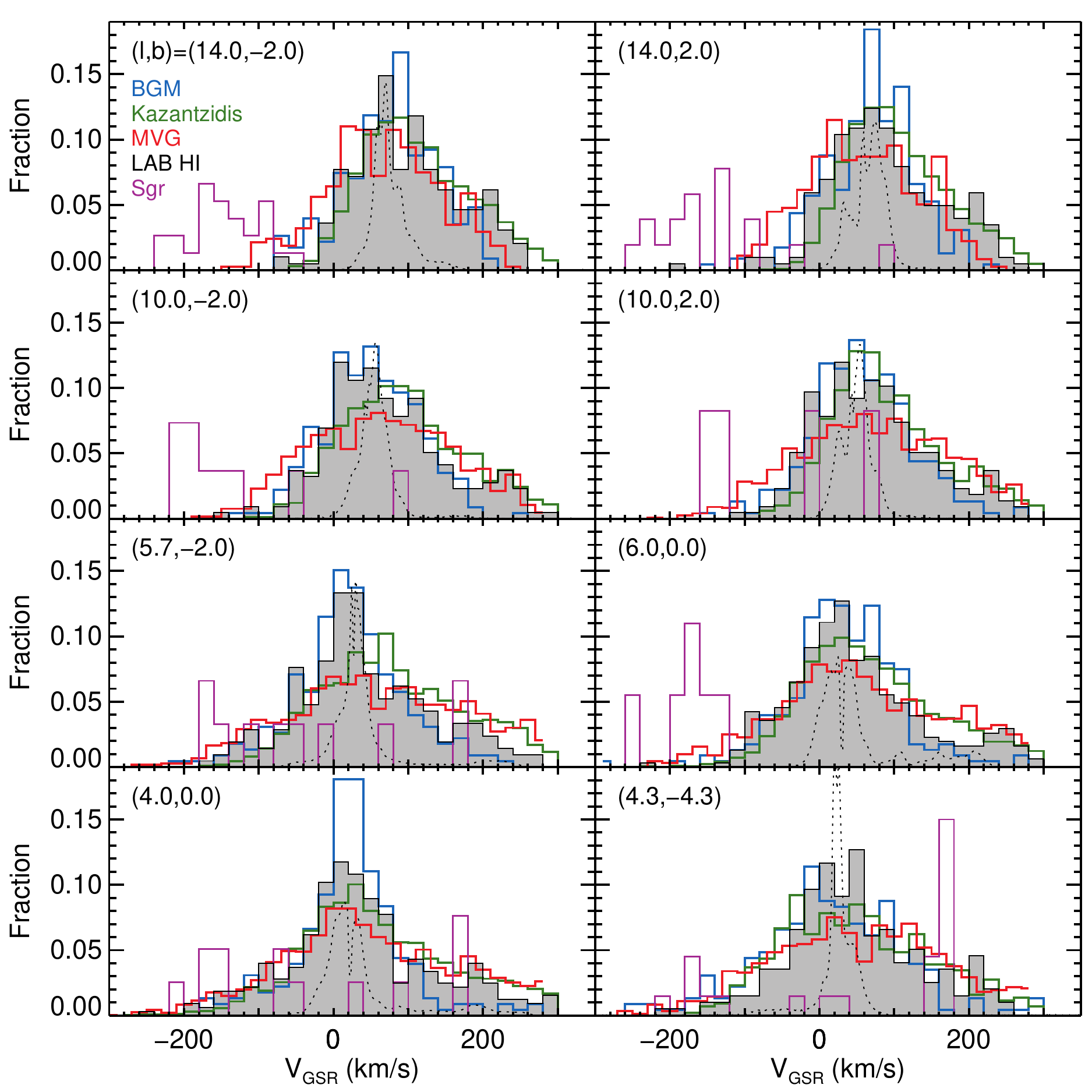}
\end{center}
\caption{Velocity ($V_{\rm GSR}$) histograms (fractions) of the APOGEE fields (black solid line and shaded region),
compared to a variety of models and other data.
The population synthesis ``Besan\c{c}on'' model of \citet{Robin03,Robin12}, which incorporates multiple 
kinematical prescriptions as described in \S \ref{subsec:Besancon} (BGM, blue).
The $N$-body model of \citet{Kazantzidis08} with initial disk scale-height of 200 pc
as described in \S \ref{subsec:Kazantzidis} (green).
The $N$-body model of \citet{Martinez-Valpuesta11} as described in \S \ref{subsec:MVG} (MVG, red).
Integrated HI data within the field's FOV from the LAB survey
\citep[][dotted black line scaled up by a factor of five]{Kalberla05}.
Predicted distribution of stars from the Sagittarius dSph streams, according to the model
of \citet[][purple line scaled down by a factor of three]{Law10}.  The Sagittarius model has been
limited to points that could possibly fall within the APOGEE sample (i.e., we assumed all points
had tip-RGB magnitudes, calculated their apparent magnitude, and removed all points too faint to
be included in the sample).
}
\label{fig_rvcomparisons}
\end{figure*}

\subsection{Comparison to Models and Other Data}
\label{subsec:modelcomparison}

Figure \ref{fig_rvcomparisons} compares the observed APOGEE RV histograms
(black solid lines and shaded region) to the above models.

The blue line shows one ``observation'' of the BGM, selected using the simulated
photometric information and APOGEE targeting algorithms with model
photometric errors and no reddening.  The BGM reproduces the main features of most fields quite well,
but not the high-velocity peak.  Multiple simulated observations
of the BGM were performed and found to be very similar,
indicating that stochastic effects were small.
Only one observation is shown for clarity.

The RV histograms of the Kazantzidis model are shown in green.  This model cannot
be mock ``observed'' like the BGM because synthetic photometry is not available.
The RV distributions are a fairly good representation of the APOGEE histograms, but
sometimes show a poorer match to the main RV peak, compared to the BGM.
This model does show significantly more
stars at high velocity and roughly match the fraction and distribution trend (decreasing
number to even higher velocities) observed in the APOGEE data.
However, it does not reproduce the ``trough'' between the main and high-velocity
peaks seen in the APOGEE data.

The MVG model (red), as with the Kazantzidis models, reproduces the
high-velocity stars much better than the BGM, although often ``over-predicting''
them.
As in the Kazantzidis model, the MVG high-velocity stars are in the bar.
This model also shows the interpeak ``trough'' in some of the fields
(e.g., [14.0,$-$2.0], [6.0,0.0], [4.0,0.0]).

The Leiden-Argentine-Bonn (LAB) \hi data \citep{Kalberla05}, shown as a dotted black line,
display high-velocity peaks in some of the fields, but often at an RV different
from the APOGEE peaks.  Gas kinematics are complicated in the inner Galaxy but 
are dominated by forces influenced by the bar \citep[e.g.,][]{Fux99}.
There does not appear to be a strong gaseous component of this new high-velocity stellar feature.

\begin{deluxetable*}{ccrccccccc}
\tablecaption{APOGEE Bulge Fields}
\tablewidth{0pt}
\setlength{\tabcolsep}{0.05in}
\tablehead{
\colhead{Field Name} & \colhead{$l_{\rm cen}$} & \colhead{$b_{\rm cen}$} & 
\colhead{Diameter} & \colhead{$\langle V_{\rm GSR} \rangle$} & \colhead{$\sigma_{\rm V}$} & 
\colhead{N$_{\rm stars}$\tablenotemark{a}}  & \colhead{$\langle V_{\rm GSR} \rangle_{\rm high}$} & \colhead{$V_{\rm thresh}$\tablenotemark{b}} &
\colhead{N$_{\rm high}$ (\%)} \\
\colhead{ } & \colhead{(deg)} & \colhead{(deg)} & 
\colhead{(deg)} & \colhead{(km s$^{-1}$)} & \colhead{(km s$^{-1}$)} & 
\colhead{ } & \colhead{(km s$^{-1}$)} & \colhead{(km s$^{-1}$)} & 
\colhead{ }
}
\startdata
004$+$00  & 4.0  & 0.0  & 1.0  & 39.4  & 100.9  & 376 & 203.6  & 141  & 56 (14.9) \\
004$-$04 & 4.3  & -4.3  & 1.0  & 47.1  & 89.8   & 246 & 214.5  & 156  & 31 (12.6) \\
006$-$02 & 5.7  & -2.0  & 1.0  & 39.5  & 88.4   & 245 & 200.0  & 146  & 29 (11.8) \\
006$+$00  & 6.0  & 0.0  & 1.0  & 50.2  & 96.1  & 364 & 234.3  & 177  & 39 (10.7) \\
010$-$02  & 10.0  & -2.0  & 1.0  & 72.0  & 80.6 & 249 & 221.7  & 170  & 28 (11.2) \\
010$+$02  & 10.0  & 2.0  & 1.0  & 69.5  & 76.2  & 247 & 233.0  & 180  & 22 (8.9) \\
014$-$02  & 14.0  & -2.0  & 2.0  & 93.7  & 77.3 & 247 & 206.8  & 176  & 32 (12.9) \\
014$+$02  & 14.0  & 2.0  & 2.0  & 85.2  & 85.2  & 251 & 195.1  & 171  & 36 (14.3) \\
\enddata
\tablenotetext{a}{Some fields have more than 300 stars due to multiple (distinct) plates being observed.}
\tablenotetext{b}{The velocity threshold for the high-velocity stars.}
\label{table_rv}
\end{deluxetable*}

Finally, because the Sagittarius dwarf spheroidal galaxy (Sgr) and part of its
tidal tails lie behind the bulge in this general direction, we test the APOGEE RVs
against the recent \citet{Law10} Sgr model (purple).  It predicts very few Sgr 
particles in our fields and at RVs quite different from the APOGEE high-velocity peaks.

\section{Discussion}
\label{sec:discussion}

We detect high-velocity peaks in our APOGEE bulge RV distributions
previously undetected in the data of BRAVA or any other survey; this might be due to the different
regions sampled by the surveys (i.e., APOGEE probing lower $|b|$). 
We compare our stellar data to various models and \hi data to discern the nature of this newly
found ``stellar population'' of the bulge.

One potential explanation of this RV feature is that these stars are part of the tidal tail of
Sgr, which lies close to the APOGEE bulge fields
and has a mean velocity of \vgsre$\sim$+170 \kms at these longitudes \citep{Law10}.
However, initial distance estimates (using preliminary APOGEE stellar parameters and
Girardi et al.\ 2002 isochrones)
of $\sim$5--10 kpc for the APOGEE stars (also see isochrones in Fig.\ \ref{fig_mapcmd}b) make this
explanation unlikely, because Sgr and its tidal tails are $\sim$29 kpc distant \citep{Siegel11}.
In addition, the \citet{Law10} model predicts very few high-velocity Sgr tidal tail stars in
our fields (Fig.\ \ref{fig_rvcomparisons}).

Could these stars be a new substructure in the MW halo?  We find this explanation unlikely because
(1) the distribution of high-velocity stars in the dereddened CMD (Fig.\ \ref{fig_mapcmd}b) is almost
identical to the rest of the stars in our fields (which should be dominated by the bulge/bar),
with rough distances of
$\sim$5--10 kpc, whereas a halo substructure should be more tightly clumped around one
distance and metallicity; (2) the high-velocity stars constitute a large fraction ($\sim$10\%)
of stars in the APOGEE fields, which is much larger than would be expected for halo substructure; and
(3) the high-velocity stars are detected in many fields covering a large range of longitude,
which would imply a truly enormous structure.

We therefore conclude that the high-RV stars are most likely members of the bar/bulge.  
Stars at these velocities and in comparable proportions of the total stellar population
are predicted by the Kazantzidis and MVG $N$-body models;
isolation and examination of these high-RV particles in the models reveal their membership in the Galactic bar.
Furthermore, these model particles are grouped together
on the ``far'' side (leading edge; see Fig.\ \ref{fig_simrvtrend}b) of the bar, suggesting
a coherent spatial/dynamical feature, 
rather than a sampling of stars with coincidentally similar velocities but on intrinsically
unassociated orbits.  (The lack of high-RV stars in the Besan\c{c}on model may simply be due to
their use of highly-smoothed kinematical prescriptions.)

The nature and significance of the trough at intermediate velocities (\vgsre$\sim$140--180 \kmse) is not clear.
We have isolated the $N$-body particles occupying the RV trough but could not
identify any group property, such as large distance, that would preferentially remove the
corresponding stars from the APOGEE sample.
It is possible that rather than an intrinsic trough (or drop in stellar surface density),
the distribution is caused by an enhancement in phase space 
at high-velocity from the combination of stars piling up at certain phases of
their orbit \citep[e.g., apogalacticon,][]{Bureau99} and projection effects.
This explanation is supported by the structure seen in the fluid-dynamical
model of \citet{Weiner99}.  In their Figures 6 and 8, a concentration of high-RV
gas exists on the leading edge of the bar, with velocities and Galactic positions consistent with the
high-RV particles ``observed'' in the Kazantzidis and MVG models.  This concentration would
correspond spatially with stars on x$_2$ orbits in the bar, perpendicular to the bar's major axis.
In this scenario, the observed RV trough simply reflects the low count of stars at high RVs
that are not in this dense dynamical structure.
Why this property is not reproduced in the $N$-body models may be due to the resolution
of the stellar structures contained within the model or the stellar density laws governing them.

The weight of evidence therefore suggests that the best explanation for the high-velocity stars is
that they are members of the MW bar, grouped on particular orbits or phases of their orbit.
The $N$-body models predict that high {\em negative} velocity stars should be seen at symmetrical
longitudes in the fourth quadrant.
Additional APOGEE data, follow-up data in the fourth quadrant, and future, more detailed
model comparisons will yield insight into the nature of these features.

\acknowledgements

We thank Stelios Kazantzidis for giving us permission to use his N-body model.
Support for D.L.N was provided by SDSS-III/APOGEE and
NSF grant AST-0807945.
G.Z. was supported by a NASA Earth \& Space Science Fellowship.
A.C.R. acknowledges support from the French Agence Nationale de la Recherche
under contract ANR-2010-BLAN-0508-01OTP. 
We thank the anonymous referee for useful comments that improved the manuscript.

Funding for SDSS-III has been provided by the Alfred P. Sloan
Foundation, the Participating Institutions, the National Science
Foundation, and the U.S. Department of Energy Office of Science. The
SDSS-III website is http://www.sdss3.org/.
SDSS-III is managed by the Astrophysical Research Consortium for the
Participating Institutions of the SDSS-III Collaboration including the
University of Arizona, the Brazilian Participation Group, Brookhaven
National Laboratory, University of Cambridge, Carnegie Mellon
University, University of Florida, the French Participation Group, the
German Participation Group, Harvard University, the Instituto de
Astrofisica de Canarias, the Michigan State/Notre Dame/JINA
Participation Group, Johns Hopkins University, Lawrence Berkeley
National Laboratory, Max Planck Institute for Astrophysics, Max Planck
Institute for Extraterrestrial Physics, New Mexico State University,
New York University, Ohio State University, Pennsylvania State
University, University of Portsmouth, Princeton University, the
Spanish Participation Group, University of Tokyo, University of Utah,
Vanderbilt University, University of Virginia, University of
Washington, and Yale University.

This publication makes use of data products from the Wide-field
Infrared Survey Explorer, which is a joint project of the University
of California, Los Angeles, and the Jet Propulsion
Laboratory/California Institute of Technology, funded by the National
Aeronautics and Space Administration.


\end{document}